# Alternative set of defining constants for redefinition of four SI units

V V Khruschov[1,2]

[1] Centre for Gravitation and Fundamental Metrology, VNIIMS, 119361 Moscow, Russia

[2] National Research Centre Kurchatov Institute, 123182 Moscow, Russia

E-mail: khkon@vniims.ru

**Abstract**

We discuss different sets of constants with fixed values for a transition to new definitions of four SI units (the kilogram, mole, ampere and kelvin). A notion of constant's order in a given system of units is suggested. Several criteria for the optimum choice of constants and an alternative set of fixed constants applicable for new definitions of the four SI units are considered. We analyse the set consisting of the Planck constant, the Avogadro constant, the Boltzmann constant and the magnetic constant.

1. **Introduction**

The existing systems of units of physical quantities are a part of the necessary toolkit of science, modern technology, industry and international trade, so as any tools, they should adequately correspond to the present-day state of the art. However the International Prototype of the Kilogram (IPK), used in the current definition of the kilogram of the SI, is the artifact made of the platinum-iridium alloy. The temporal instability of the IPK have been revealed on the level of $5 \times 10^{-10}$ kg per year [1] (see also [2]). It is unacceptable for high precision measurements and long-term storing of obtained results. Redefinition of the four SI base units, being now under preparation [3, 4], can be considered as a response to this challenge. The planned SI revision grounds on the proposal to define the base units by fixing exact values of corresponding physical constants (PC), following the principle that was already used in the definitions of the ampere in 1946 and the metre in 1983 [5, 6, 7]. In the SI Brochure 9[th] edition Draft [4] the new term is used instead of fixed PC, namely, *defining constants* (DC). The main obstacle for adoption of new definitions, as soon as it had been realized, was the insufficient precision of experimental DC values.

The proposal for redefinition of a number of SI units was published in 1999 [8] and the complete set of proposals was published in 2006 [6]. Several DC sets were considered.

In particular it was suggested to fix without any uncertainty the values of the Planck constant ($h$), elementary charge ($e$), Boltzmann constant ($k$) and Avogadro constant ($N_A$) for redefinition of the kilogram, ampere, kelvin and mole [4, 6, 7]. This DC set is considered as the most preferable now. For instance, in the SI Brochure 9[th] edition Draft the presentation of SI with new definitions of four base units (the New SI or the Modified SI (MSI)) is made on this assumption.

The current situation with a transition to the MSI is reported in Resolution 1 of the 24th CGPM in 2011, where it was proposed "*to continue work towards improved formulations for the definitions of the SI base units in terms of fundamental constants, having as far as possible a more easily understandable description for users in general, consistent with scientific rigour and clarity*". The 25th CGPM in 2014 pointing to the progress from the 24th CGPM nevertheless took into account the insufficient precision of measured DC values and recommended to continue the execution of Resolution 1 of the 24th CGPM. It is supposed the 26th CGPM will adopt the MSI in 2018 [3].

The purpose of this paper is to discuss critically proposed new definitions of the four SI units in the light of the current theoretical and experimental results [9-13]. We introduce a new notion, namely, a constant's order in a given system of units and suggest a list of criteria for a choice of the appropriate DC set. An alternative DC set applicable for new definitions of the four SI units is suggested as well. The set consists of $h$, $k$, $N_A$ and the magnetic constant $\mu_0$ (System D). Advantages of this set are compared with those of the most preferable now set with $h$, $e$, $k$ and $N_A$. We expect such consideration is useful for a sensible and relevant choice of the optimum DC set.

The paper is organized as follows. Section 2 suggests a new DC classification according to their dimensional relationship to the base units ('the DC order') and discusses the criteria for a preferable choice of a DC set for new definitions of the SI units. Section 3 analyses different DC sets proposed for a redefinition of the four SI units. In Section 4, new definitions of the kilogram, mole, ampere and kelvin are considered on the basis of fixation of the $h$, $\mu_0$, $k$ and $N_A$ values. Section 5 is devoted to application of the notion of the DC order and the considered criteria to a choice of a DC set. It is mentioned advantages and disadvantages of System D relative to other sets. The conclusion summarizes our proposals for new definitions of the four SI units.

2. **DC classification and criteria for a preferable choice of DC**

In the SI Brochure 9[th] edition Draft [3] the presentation of the MSI is made on the basis of $h$, $e$, $k$ and $N_A$ fixed values. The general DC notion is used instead of more specific notions such as PC, a constant of nature, the IPK, the material parameter and so on. The totality of constants are divided in the four classes, namely, fundamental constants of nature, special atomic parameters, conversion factors, and technical constants. However this classification has some shortcomings. For instance, the speed of light in vacuum $c$ is a fundamental constant of nature, but it becomes at the same time a conversion factor after fixing its value. We use the term "conversion factor" which uses in the framework of the New SI [4]. $N_A$ also becomes the conversion factor between the unit for amount of substance and the unit for counting entities (unit 1) [4] (see also [14]), but it can be considered as the constant of nature (Avogadro's law). It is known that the number of constants increases or decreases along with the number of the base units [15]. Besides that, the number of constants in the framework of the definite system of units depends on an used physical theory or a model. For example, there are not the $c$ and $h$ constants in the frame of the nonrelativistic classical mechanics, but they are in the relativistic quantum mechanics. In [16] a vector space was associated with dimension powers of units and a test of whether a set of constants forms a complete set was proposed. Different classifications of PCs are suggested taking into account their role in description of physical phenomena (see, e.g. [17]).

Here, we propose a new specification of constants that is connected with the base units of any given system of units. The constant which does not depend on the base units of the system is named constant of *order zero*. A constant which depends on a base unit is named constant of *order one* and so on. In Sec. 5 we apply the classification of constants according their orders for a choice of a DC set.

At first glance a PC value cannot be fixed by virtue of the fact that it contains both a *estimated value* and a *uncertainty* associated with the value. However the PC value can be fixed for the redefined unit if this does not increase to a sizeable extent the uncertainty of the realization of the unit and gives additional benefits. Such a fixed PC will be the DC for the considered unit. It is desirable that a DC set be consistent and minimal, with the aim of any DC using do not lead to a contradiction with using of other DCs. The status of any DC is not absolute and can change due to generalization of a used theory or increasing of precision of some DCs.

Anyway, the key requirement to the MSI is that it should not aggravate the situation in any respect for users as compared with the current SI. This means a necessary continuity with respect to the current SI, which implies establishing *the same set of base units and the same values of the units as they exist by the day of revision* in order that the whole enormous set of the existing measurement data could be preserved without correction.

Evidently, the revision of the SI should not also worsen the stability properties of realization of units against the old ones. For instance, new BIPM prototypes of the mass unit should have a confirmed temporal instability not worse than $5\times10^{-10}$ per year, and this requirement should hold for any realization of the kilogram at any time and place. In this respect, it is appropriate to recall that the detected temporal instability of the IPK copies for 100 years [1, 2] must be ended by the planned transition to the New SI.

For concordance between the system of DCs and physical theories, it is desirable that the number of fixed constants should be minimum possible, and the relation between the DC and the corresponding unit should be as simple as possible. The same conditions are desirable for successful teaching of the fundamentals of metrology at universities and colleges. This goal is achieved if the base unit and the corresponding DC have the same physical dimension or their dimensions relate each other as simple as possible. For example, the dimensions of the velocity of light and the metre, the electron charge and the ampere differ only by a certain power of time. The dimension of the kilogram is the same as the dimension of the unified atomic mass unit $m_u$ (= (mass $^{12}$C)/12). In terms of constants of different orders having introduced above the criterion needed for this purpose can be formulated as follows: *the DC order should be as low as possible*.

Thus we suggest the following criteria for choosing the optimal DC set for new definitions of SI units (see also [18 - 20]): a) continuity between the old and new definitions, b) the stability requirement for transfers of the unit values, c) a minimum of DCs, d) use of DCs with the minimal possible orders.

### 3. Analysis of DC sets proposed for redefinition of the four base SI units

At the present time discussions about strengths and shortcomings of different new definitions of base SI units still continue (see, e.g. [9, 19, 20-22]). The preferred variant is the DC set consisted of $h$, $e$, $k$, and $N_A$ [6]. Other variants have been proposed [8, 9, 19, 20]. For instance, the opinion of the authors of Ref. [9] consists in the fact that the level of precision of the CODATA-2014 data set [13] provides a different context for the choice of a optimal DC set than was possible in 2007 at the 23th CGPM [6, 23]. Five possible variants of the SI with different DC sets (including the present SI) were considered in the work [9]. For instance, in the present SI the definitions of the kilogram, the ampere, the kelvin and the mole are based on the fixed values of the IPK mass ($m(K)$), $\mu_0$, the triple point of water ($T_{tpw}$) and the molar mass of $^{12}$C ($M(^{12}$C)). In the work [9] the present SI is named as System A. Remaining variants of the MSI are based on the fixed values of different sets containing $h$, $e$, $k$, $N_A$, $m_u$, $\mu_0$. Since in the frameworks of all variants $k$ and $N_A$ are fixed then we can distinguish these variants picking two DC of $h$, $e$, $m_u$, $\mu_0$. So System B contains $h$, $e$, System C contains $e$, $m_u$, System D contains $h$, $\mu_0$, System E contains $m_u$, $\mu_0$. Let us note, that the pair $h$, $m_u$ is excluded in ref. [9] because fixing the numerical values of both would redefine the second and the pair $e$, $\mu_0$ would add no benefits over the B, C, D and E variants.

Let us consider in more detail Systems B, C, and E. System D is considered in Sec. 4. System B is the variant of MSI proposed in the work [6] and noted in Resolution 1 of the 24 CGPM. System B provides zero uncertainty for the Josephson constant $K_J$ and the von Klitzing constant $R_K$, which are used as practical electromagnetic standards, i.e. this variant settles the current problem of electrical metrology, namely, the existence of the fixed $K_J$ and $R_K$ values accepted in 1990. In 1990 the $K_J$ and $R_K$ values conform to the experimental ones within their uncertainties, but now it is not so. Thus the 1990 values should be abrogated and the $K_J$ and $R_K$ values should be fixed on the base of the best fit CODATA values of $h$ and $e$ assuming the usual relations for $K_J$ and $R_K$.

Notice that for System B the constants $\mu_0$ and $M_u$ (the molar mass constant) must change depending on experimental values and uncertainties. For instance, this dependence for $\mu_0$ can be parameterized as $4\pi \times (1+\delta) \times 10^{-7}$ N/A$^2$, where the value of $\delta = (\alpha/\alpha_{2018} -1)$, $\alpha_{2018}$ being the best available experimental value of $\alpha$ at the time of redefinition [9]. $u_r$ of $\delta$ is the same as $u_r$ of $\alpha$, taking the $\alpha_{2018}$ value to be exact. So if deviations of $\mu_0$ and $M_u$ from the values $4\pi \times 10^{-7}$ N/A$^2$ and 1 g mol$^{-1}$ will be small enough then we shall have no problems "in practice". However, the factor $(1+\delta)$ is difficult to explain when $\mu_0$ needs to be introduced in textbooks in electromagnetism or in educational process and it is not compatible with Resolution 1 of the 24th CGPM that the New SI should be "easily understandable for users in general, consistent with scientific rigour and clarity" [3]. This is similar to the problem with the varied molar mass constant $M_u$ in the framework of Systems B and D.

New definitions of the kilogram and the mole in Systems E and C can be realized with the help of both the watt balance and the silicon sphere devices. Equivalence of these methods follows from the known relation: $N_A h = A_r(e) M_u c \alpha^2 / (2 R_\infty)$, where $M_u = M(^{12}C)/12$, $M_u = N_A m_u$ coincides with $M_{u0} = 1$ g mol$^{-1}$ in the present SI, $A_r(e)$ is the relative atomic mass of the electron ($=m_e/m_u$), $R_\infty$ is the Rydberg constant. The value of the molar Planck constant is known with $u_r = 4,5 \times 10^{-10}$ [13], so there is not any loss of precision at determination of the $N_A$ value with the help of $h$ value and vise versa. The systems E and C have advantages for definitions of the kilogram and the mole [24] which did not early take into account due to large uncertainties for $R_K$ and $K_J$ leading to degradation of electromagnetic measurements accuracy. System C is considered in detail in the works [20, 22, 25, 26]. This system is intermediate between System B [6] and System E [9], and is optimal with respect to the criteria presented in Sec. 3 (see also Sec. 5).

### 4. New definitions of the kilogram, ampere, kelvin and mole in System D

Each version of a DC set has its advantages and disadvantages, which is desirable to consider and discuss at a preparation of the 26 CGPM decision. In the present paper an alternative set of fixed constants applicable for new definitions of four SI units is considered in detail. This set (System D) consists of $h$, $\mu_0$, $k$, and $N_A$ (see below and section 5) and has been proposed for the first time by the working group on base units and fundamental constants of the *Académie des Sciences*, Paris (it was proposed to fix the Planck charge $q_P = (2\varepsilon_0 hc)^{1/2}$) [27].

In System D the new definitions of the mole and the kelvin are based on fixed values of $N_A$ and $k$ as well as in Systems B, C and E. The kilogram is based on a fixed value of $h$ as in System B, and the ampere is based on the fixed value of $\mu_0$ as in the present SI. By reason of the non-fixed value of the electron charge $e$ in System D we have experimental uncertainties for $R_K$ and $K_J$, however they are very small (~ $10^{-10}$) and negligible. Notice, it is only needed for precise electromagnetic measurements that the $R_K$ and $K_J$ values be stable and known with uncertainties smaller than some limits. The fine structure constant $\alpha$ plays a central role in determining the uncertainties of $R_K$ and $K_J$. Recently the value of $\alpha$ has been confirmed by the completely independent, non QED, route of atomic recoil experiment (e.g. measurements of $h/m(^{87}\text{Rb})$ [28]) with low uncertainty. Improvements in $\alpha$ determinations will continue that will give us the high accuracy of determination of the $R_K$ and $K_J$ constants.

Let us draw attention to the advantage of keeping the fixed $\mu_0$ value in System D, $\mu_0 = 4\pi \times 10^{-7}$ N/A$^2$. First and foremost it is consistent with the fixation of $c$ in 1983. In so far as the dielectric and magnetic permittivity of free space $\varepsilon_0$ and $\mu_0$ obey the relation: $\varepsilon_0 \mu_0 c^2 = 1$, then $\varepsilon_0$ must be fixed too and be equal to the SI value: $\varepsilon_0 = 8.854187817 \ldots$ F m$^{-1}$. So, taking into account $\alpha = e^2/(2\varepsilon_0 h c)$ we have $u_r(e^2) = u_r(\alpha)$. Afterwards for estimations of different constants' uncertainties in System D one may note that the results of spectroscopic measurements depend on the metre and the second only, which are affected by the choice of DC. So, for instance, it is sufficient to represent $m_u$ and $e$ through the fixed constants $h$ and $\mu_0$ and the measured constants: $\alpha$, $A_r(e)$ and $R_\infty$. Fixed $\mu_0$ implies that the vacuum impedance $Z_0$ is also fixed. So one can determine not only $R_K$ by direct comparison with $Z_0$ by means of the calculable capacitor experiment, but also $K_J$ with the watt balance, without depending upon the veracity of the formulae connecting these constants to $e$, $h$ and $\alpha$ [27]. Thus System D is best suited to lead to progress in science and modern technology.

5. **Orders of DCs and application of criteria to Systems B, C, D and E**

Let us find orders of DCs considered for redefinition of the four base SI units such as $h$, $e$, $k$, $N_A$, $m_u$ and $\mu_0$. It is needed in order to apply the list of criteria identified for choosing constants in Sec. 3. The dimensions of these constants with respect to the base SI quantities are as follows

$$[h] = ML^2T^{-1}, \quad [e] = TI, \quad [N_A] = N^{-1},$$
$$[k] = ML^2T^{-2}\theta^{-1}, \quad [m_u] = M, \quad [\mu_0] = LMT^{-2}I^{-2}.$$

We have the following orders of these constants according to the definition of the constant's order given in Sec. 2.

$$O(h) = 3, \ O(e) = 2, \ O(N_A) = 1, \ O(k) = 4, \ O(m_u) = 1, \ O(\mu_0) = 4,$$

where $O(DC)$ stands for the order of DC. Moreover one can define the order of a system of constants $S = \{c_1, c_2, c_3, \ldots\}$:

$$O(S) = \sum_i O(c_i).$$

So $O(B) = 10$, $O(C) = 8$, $O(D) = 12$, $O(E) = 10$.

In Systems B, C, D and E we use $k$ and $N_A$ for new definitions of the temperature unit and the amount of substance unit correspondingly. There is no other alternative but only the $k$ and $N_A$ constants are suitable for this purpose. It is remained to choose two constants for new definitions of the mass and current units. If one directly uses the *d* criterion written in Sec. 2 then the $m_u$ and $e$ constants will be obtained. Thus System C, which has been considered in the work [8], must be chosen and has been offered for the redefinition of the kilogram, mole, ampere and kelvin in the works [20, 22, 25, 26]. The realization of these new definitions must be carried out with account of the *b* criterion from Sec. 2.

Above we uniquely determine System C for the redefinition of the four base SI units with the help of the DCs orders and the criteria considered in Sec. 2. However, there are other criteria that can be used for choosing a DCs set. For instance, the authors of the draft 9[th] edition of the SI Brochure [4] have chosen System B barely taking into account that any coherent unit of the SI can be expressed in terms of the DCs. System B provides great advantages for the electromagnetic metrology and it is the important reason to choose the $h$, $e$, $k$ and $N_A$ constants. But other sets still persist which have their own merits. In Sec. 4 we propose system D based on the $\{h, \mu_0, k, N_A\}$ set. The advantages of this set are the stability of the $h$ and $\mu_0$ values and the fixation of properties of the electromagnetic vacuum in which quantum phenomena occur. The fixation of the $e$ and $m_u$ values leads to the fixation of electromagnetic and mass properties for atomic particles. We have two systems (B and C) containing $e$ and two systems (D and E) containing $\mu_0$. It is known that the electrical charge value varies with a interaction distance or a energy scale of the process, for example, $\alpha^{-1} \sim 137$ appropriate at very low energy, while $\alpha^{-1} \sim 128$ at the scale of the Z boson mass [28]. So it is preferable to use systems D or E for new definitions of the base SI units from this point of view.

## 6. Conclusions

We have analyzed the suggested versions of new definitions of the kilogram, ampere, kelvin and mole based on fixing the exact values of certain DCs, and formulated the notion of the constant's order in the given system of units. Following the results of the works [9, 19] we have considered the four versions of the New SI (systems B, C, D and E) taking into account progress in experimental determinations of DCs values. New experimental results for the Planck and Avogadro constants [10-12], as well as the CODATA 2014 adjustment of PC, certainly, will have an influence on a choice of four possible formulations of the base SI units. Thus the arguments for and against different formulations should be changed since they were first proposed in 2006.

In the present paper the new classification of DCs based on the notion of the order of DC in the given system of units is introduced and some criteria are formulated for choosing the set of DCs suitable for redefinition of the base SI units. The New SI version with fixed $h$, $\mu_0$, $k$, $N_A$ (System D) and its advantages are considered in detail. We can say that fixing of $h$ and $\mu_0$ leads to some fixation of space-time properties associated with quantum phenomena. It is consistent with the fixation of $c$ in 1983 and gives the fixation of $\varepsilon_0$, as well as the fixation of the vacuum impedance $Z_0$.

When System D is compared with System B, it is apparent that System B has advantage for electromagnetic measurements because the practical units $\Omega_{90}$ and $V_{90}$ (or, more precisely, $\Omega_{2018}$ and $V_{2018}$) will turn to the New SI units. However, using the fixed $e$ value is the drawback due to its dependence on the energy scale of the process and possible space and time variations of $\alpha$ [29]. The practice of using fixed values of $K_J$ and $R_K$ can be preserved at a certain accuracy level in framework of Systems D, C and E together with further measuring $e$ or $\alpha$ more and more accurately for testing the existing theories and a search for new ones. From our point of view System D is the best choice for progress in science and modern technology notwithstanding that the CCU rejected this proposal [30].

The aim adopted about ten years ago, i.e. the determination of precise values of the Planck and Avogadro constants concordant with each other, had been achieved by 2015, when these constants had been measured with $u_r = 2\times10^{-8}$ by the silicon sphere method [11] and the watt balance method [10, 12]. Thus, the results of the recent NMIJ, BIPM, PTB, INRIM, NIST and NRC experiments [10-12] allow to achieve the $10^{-8}$ level demanded by the 24th and 25th CGPM decisions [3] and, most likely, will bring to the adoption of new definitions of the kilogram and mole in 2018. There is yet time to consider the advantages and shortcomings of the suggested new definitions and their different versions taking into account the fulfillment of appropriate criteria and further progress in precision of relevant DC measurements' results.

**Acknowledgments**

The author would like to thank the referees for the constructive criticism that allows improving of the text.